\documentclass[aps,prl,a4paper,twocolumn,showpacs]{revtex4}
\usepackage{graphicx}

\begin{document}

\title{ Increasing d-wave superconductivity  by on site repulsion  }

\author{E. Plekhanov$^{a}$, S. Sorella$^{a,c}$, and M. Fabrizio$^{a,b}$}

\affiliation{$^a$ Istituto Nazionale di Fisica della Materia and
International School for Advanced Studies (SISSA)\\
Via Beirut 2-4, I-34014 Trieste, Italy}
\affiliation{$^b$ International Center for Theoretical Physics (ICTP)\\
Strada Costiera 11, I-34100 Trieste, Italy}

\affiliation{$^c$ INFM - DEMOCRITOS, National Simulation Center\\
Via Beirut 2-4, I-34014 Trieste, Italy}

\date{27 September 2002}

\begin{abstract}
We study by Variational Monte Carlo an extended Hubbard model
away from half filled band density which contains two competing
nearest-neighbor interactions: a superexchange $J$ favoring
$d$-wave superconductivity and a repulsion $V$ opposing against
it. We find that the on-site repulsion $U$ effectively enhances
the strength of $J$ meanwhile suppressing that of $V$, thus
favoring superconductivity. This result shows that attractions
which do not involve charge fluctuations are very well equipped
against strong electron-electron repulsion so much to get
advantage from it.

\end{abstract}

\maketitle

The interplay between strong correlation and superconductivity
is one of the major problems raised by the discovery of cuprate
High-T$_c$ materials.

Indeed, within the conventional BCS theory for phonon-mediated
weak coupling superconductivity, a strong electron-electron short
range repulsion, parametrized by an Hubbard $U$, can only depress
T$_c$. Landau Fermi liquid theory identifies two main sources for
this reduction; namely, as $U$ increases, the quasiparticle wave
function renormalization $Z\leq 1$ diminishes, meanwhile the
effective mass $m_*$ increases. Thus each interaction amplitude,
including the phonon-mediated attraction $g$, acquires a
renormalization factor $Z^2$ times a vertex renormalization. If
the latter is negligible, the bare amplitude is reduced to
$g_*=Z^2 g$.  Therefore, as $U$ increases, the dimensionless
coupling which parametrizes the {\sl bare} attraction, $\lambda_0
= \rho_0 |g|$, being $\rho_0$ the uncorrelated density of states
at the Fermi level, is renormalized into
\[
\lambda_* = \rho_* |g_*| =
Z^2 \frac{m_*}{m} \lambda_0  < \lambda_0,
\]
being usually $Z\leq m/m_*$.
On the other hand the Coulomb pseudopotential  $\mu_*$ increases,
so that $\lambda_* -\mu_*$ diminishes even more pushing T$_c$ down.

By solving a model for alkali doped fullerenes
within dynamical mean field theory, it has been
recently argued\cite{Capone} that
there exist attractive channels for which vertex corrections may compensate
the wave-function renormalization factor leading to
\[
\lambda_0 \to \frac{m_*}{m} \lambda_0  > \lambda_0,
\]
which may indeed lead to an enhancement of T$_c$ by
increasing $U$. The main ingredient
was identified into a pairing mechanism not involving
the charge density operator, which is mostly subject to the renormalization
induced by $U$, but other internal degrees of freedom, like
the spin (or the orbital index, if orbital degeneracy is present),
unveiling a kind of spin-charge separation even within Landau Fermi liquid
theory.

This proposal is not far in spirit from the original Resonating
Valence Bond (RVB) scenario for High-T$_c$ superconductivity in
the $t$-$J$ model\cite{PWA}. There superconductivity occurs
naturally upon doping since the parent insulating state is well
described by an RVB state: the spin-singlet valence-bond pairs
naturally evolve into Cooper pairs. They can propagate around the
lattice only through the empty sites left behind by the holes.
This constraint easily explains a superfluid density proportional
to the hole doping. Moreover, although at small doping
superconductivity is suppressed, the energy scale associated to
the binding energy of the valence-bond pairs remains finite,
which is advocated to explain the experimentally observed
pseudogap phase of High-T$_c$ materials\cite{Randeria,Sandro}.

Within the Fermi liquid framework, the constraint of no double
occupancy appears to renormalize the quasiparticle hopping to a
value $Zt\simeq 2\delta t$, $\delta$ being the doping, while
leaving untouched the quasiparticle attraction, here provided by
the superexchange $J$. The superconducting phase of the $t$-$J$
model can be approached either from the half-filled
antiferromagnetic Mott insulator upon increasing doping or at
finite doping by decreasing temperature. In both cases, even
though the $T=0$ superconductor might still be described in terms
of Landau-Bogoliubov quasiparticles, in the RVB language
spinon-holon confined objects, the relevant excitations above
T$_c$ or in the close vicinity to the half-filled antiferromagnet
do not necessarily look as conventional quasiparticles.

For this reason, in this work we shall try to understand whether
this strongly correlated $d$-wave superconductor
can be approached at zero temperature starting from a weakly
correlated regime, where standard many-body
techniques and the well established Landau Fermi liquid theory should apply.

We consider an extended Hubbard model in
two-dimensions for the average number of electrons per site $1-\delta<1$, 
namely away from half-filling,

\begin{eqnarray}
\hat{H} &=& -t\sum_{<ij>}\sum_\sigma \left(c^\dagger_{i\sigma}
c^{\phantom{\dagger}}_{j\sigma} + H.c. \right)\,
+ U\sum_i n_{i\uparrow}n_{i\downarrow} \nonumber \\
&& + J\sum_{<ij>} \left(
\vec{S}_i\,\cdot\, \vec{S}_j -\frac{1}{4}n_i n_j \right)
+ V\sum_{<ij>} n_i n_j,
\label{Hamiltonian}
\end{eqnarray}

where, in addition to the on-site repulsion, we add a nearest
neighbor spin-exchange and a charge-repulsion term, with
strengths $J$ and $V$, respectively. These nearest-neighbor
interactions compete, $J$ favoring a $d$-wave singlet pairing
away from half-filling while $V$ opposing against it. For $V=0$
and $U$ strictly equal to $\infty$, (\ref{Hamiltonian}) reduces
to the standard $t$-$J$ model, which also corresponds to the
large $U$ limit of the pure Hubbard model, in which case $J\to
4t^2/U$. However, contrary to the latter, model
(\ref{Hamiltonian}) for $J>V$ is undoubtedly a $d$-wave
superconductor at weak coupling ($U$, $V$ and $J$ all much smaller
than $t$) also within the Hartree-Fock approximation. For this
reason model (\ref{Hamiltonian}) is more suitable to address the
issue of electron-electron correlation effects on $d$-wave
superconductivity. Moreover, since $V$ involves charge-density
while $J$ spin-density operators, the presence of both gives the
opportunity to test if $U$ indeed induces different
renormalization factors on charge with respect to spin vertices.

A variational approach which was shown to correctly reproduce
both the weak \cite{shiba} and
the strong \cite{Sorella} coupling limits of the Hubbard model
appears well suited  for model (\ref{Hamiltonian}) too.
It consists in searching by a variational
Monte Carlo (VMC) technique
for the best wave function of the form
\begin{equation}
|\Psi \rangle = A\hat{P}_{{\rm N}}\hat{P}_{{\rm
Jastrow}}\hat{P}_{{\rm G}} |\Psi_{BCS}\rangle
\label{variational-wf}
\end{equation}
where $A$ is a normalization factor, $|\Psi_{BCS}\rangle$ a BCS
wave-function \cite{notesdw}  projected by $\hat{P}_{{\rm N}}$ onto a fixed
number of particles, with a $d$-wave gap-function $\Delta_k =
\Delta_{var} (\cos k_x - \cos k_y)$, being $\Delta_{var}$ a variational
parameter. $\hat{P}_{{\rm G}}$ is a
Gutzwiller projector:
\begin{equation}
\hat{P}_{{\rm G}}=\prod_{n} \left(1 - \alpha_0
n_{n,\uparrow}n_{n,\downarrow}\right),
\end{equation}
 whereas $\hat{P}_{{\rm Jastrow}}$ a long-range Jastrow factor which
enforces the correct long-wavelength behavior of the density structure
factor:
\begin{equation}
\hat{P}_{{\rm Jastrow}} = e^{-\alpha_1\sum_{<ij>}n_i n_j
-\alpha_2\sum_{<<ij>>}n_i n_j - \ldots},
\label{jastrow}
\end{equation}
where "$\ldots$" stands for the summation over next, next-next, etc.
nearest neighbor sites (i.e. all those possible on a finite size sample).

The method is based on the Stochastic Reconfiguration (SR)
technique \cite{Sorella}, which allows to minimize the energy of a
variational wave function containing even a large number of
parameters.

To get further insight from the numerics, we compare the results
with those obtained by the Gutzwiller
Approximation (GA) \cite{gutz,gebh} for the variational wave-function without
both the long-range Jastrow factor and the projector onto a fixed
number of particles.

\begin{figure*}
\centerline{
\includegraphics[width=7.0cm]{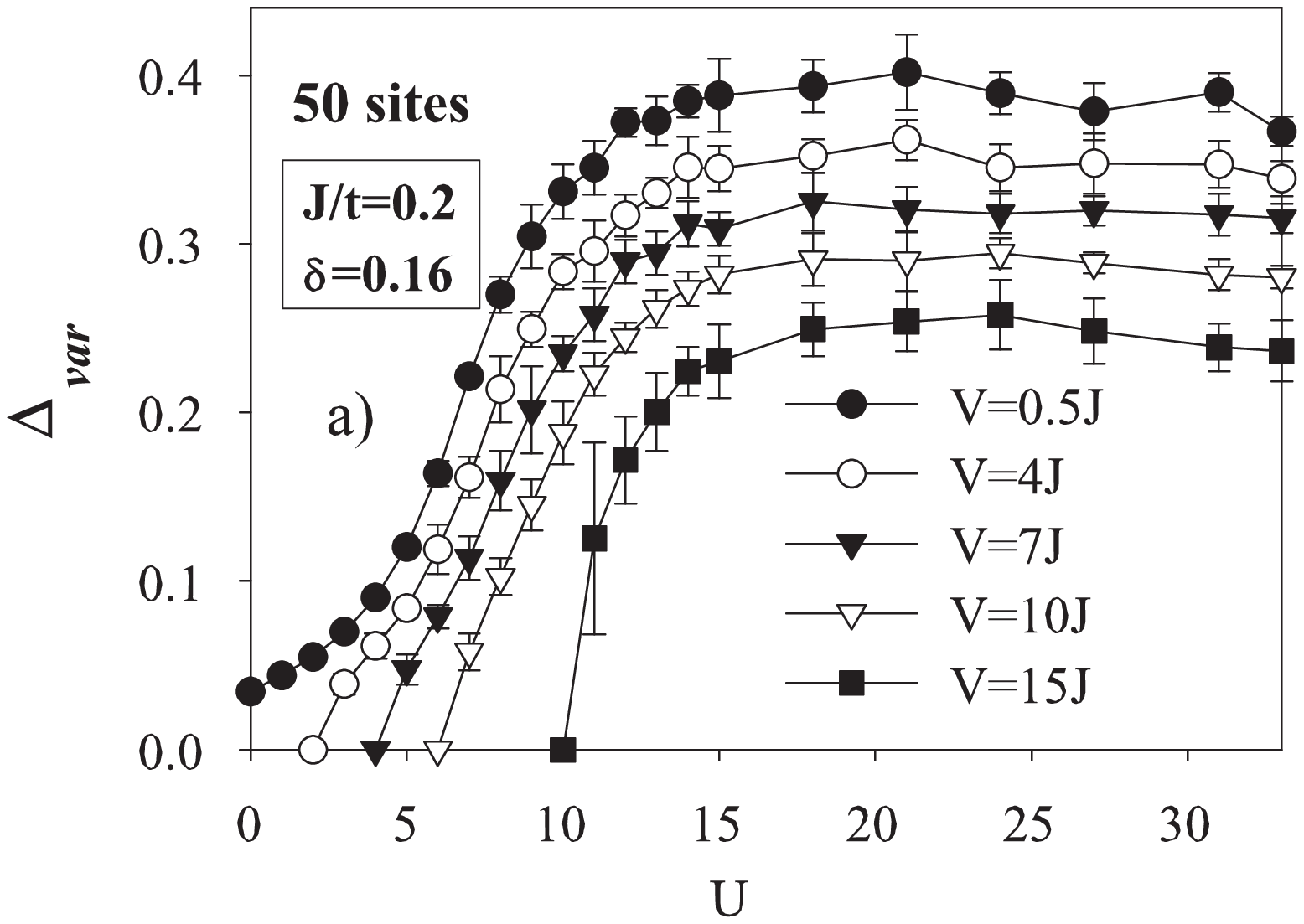}
\includegraphics[width=7.0cm]{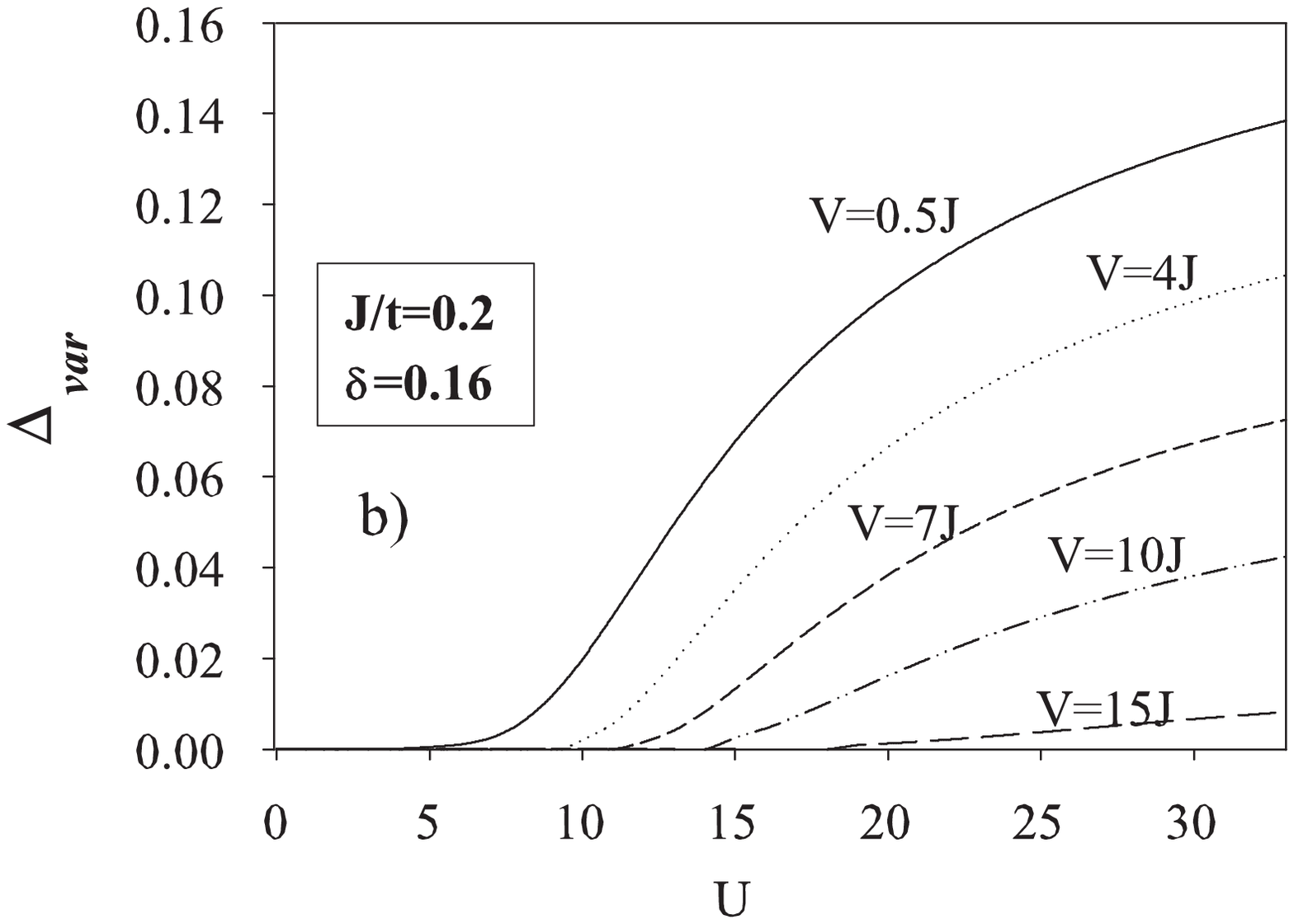}}
\caption{\label{deltas} Variational gap as function of $U$ for
different values of $V$ within variational Monte Carlo a) and GA
b).}
\end{figure*}

In Fig.~\ref{deltas} we plot the variational parameter
$\Delta_{var}$ as a function of $U$ for $J=0.2t$, $\delta=0.16$
and for different values of $V$. For $J>V$, $\Delta_{var}$ starts
finite at $U=0$ and increases with $U$. For
$V>J$, $\Delta_{var}=0$ at small $U$, in agreement with the
Hartree-Fock results. More remarkably above a critical $U_c$ a finite
$\Delta_{var}$ appears. Namely, the normal metal at $V>J$ turns
into a superconductor by increasing the on-site repulsion. Both
results can be explained within the Fermi liquid picture provided
by the Gutzwiller approximation, according to which the effective
$J_*$ acting between the quasiparticles stays essentially
unrenormalized when $U$ increases, contrary to the effective $V_*$, which is
substantially suppressed with respect to its bare value $V$.
Therefore, as $U$ increases for $J>V$, the quasiparticle
bandwidth gets reduced, the attraction staying unrenormalized, so
that the dimensionless coupling $\lambda_*$ increases, hence
$\Delta_{var}$. If $J<V$, a normal metal is stable until $V_* >
J_*\simeq J$, after which superconductivity gets in. In our
numerical study we have found that the inclusion of the long
range Jastrow factor (\ref{jastrow}) considerably improves the
simpler Gutzwiller wave function and allows larger values of
$\Delta_{var}$. However, as shown in Fig. \ref{deltas}b the
qualitative behavior of $\Delta_{var}$ vs. $U$ is reproduced
already by GA.

 Within the GA it is possible to study explicitly the
competing influence  of both $J$ and $V$ on
superconductivity. Let us consider the superconducting
contributions to the uncorrelated expectation values for nearest
neighbor sites $i$ and $j$,
\begin{eqnarray}
\langle \Psi_{BCS} | n_i n_j | \Psi_{BCS} \rangle_{SC}
&=& 2 \Delta_{SC}^2 \label{V}\\
\langle \Psi_{BCS} | \vec{S}_i\cdot \vec{S}_j | \Psi_{BCS} \rangle_{SC}
&=& - \frac{3}{2} \Delta_{SC}^2 \label{J},
\end{eqnarray}
where $\Delta_{SC} = |\langle \Psi_{BCS} | c^\dagger_{i\sigma}
c^\dagger_{j-\sigma}| \Psi_{BCS} \rangle |$ is the order
parameter of the uncorrelated wave function. In the case of
Eq.(\ref{V}), this term derives from configurations in which $i$
and $j$ are both singly occupied, both doubly occupied or one
singly and the other doubly occupied, with weights $\delta^2$,
$(1-\delta)^2$ and $2\delta(1-\delta)$, respectively, where
$\delta$ is the doping. On the contrary, Eq.(\ref{J}) has
contribution only by configurations where both sites are singly
occupied. In the limit of very large $U$, the configurations with
doubly occupied sites are projected out, hence only the
contributions from singly occupied sites survive in Eqs.
(\ref{V}) and (\ref{J}). This implies that (\ref{V}) gets a
reduction factor $\delta^2$ relatively to (\ref{J}), so that the
actual condition for superconductivity at $U\to\infty$ reads
approximately
\[
\frac{3}{4}J > \delta^2 \left(V-\frac{1}{4}J\right).
\]

Since, in the same limit, the wave function renormalization
$Z\simeq 2\delta$, we indeed recover the desired Fermi liquid
result that interactions involving the charge density operators
get renormalized down by a factor $Z^2$ with respect to those
involving spin operators. The above discussion also shows that
not all the pairing mechanisms are equally equipped against
on-site repulsion. Indeed a weak coupling $d$-wave
superconductivity might be stabilized even by a negative $V$ at
$J=0$: an explicit attraction  between charges. However, for
increasing $U$, the effective strength of this attraction would
decrease as $Z^2$ so that $\lambda_* \sim Z \lambda_0$, hence
$T_c$ would go down.

\begin{figure*}
\centerline{
\includegraphics[width=7.0cm]{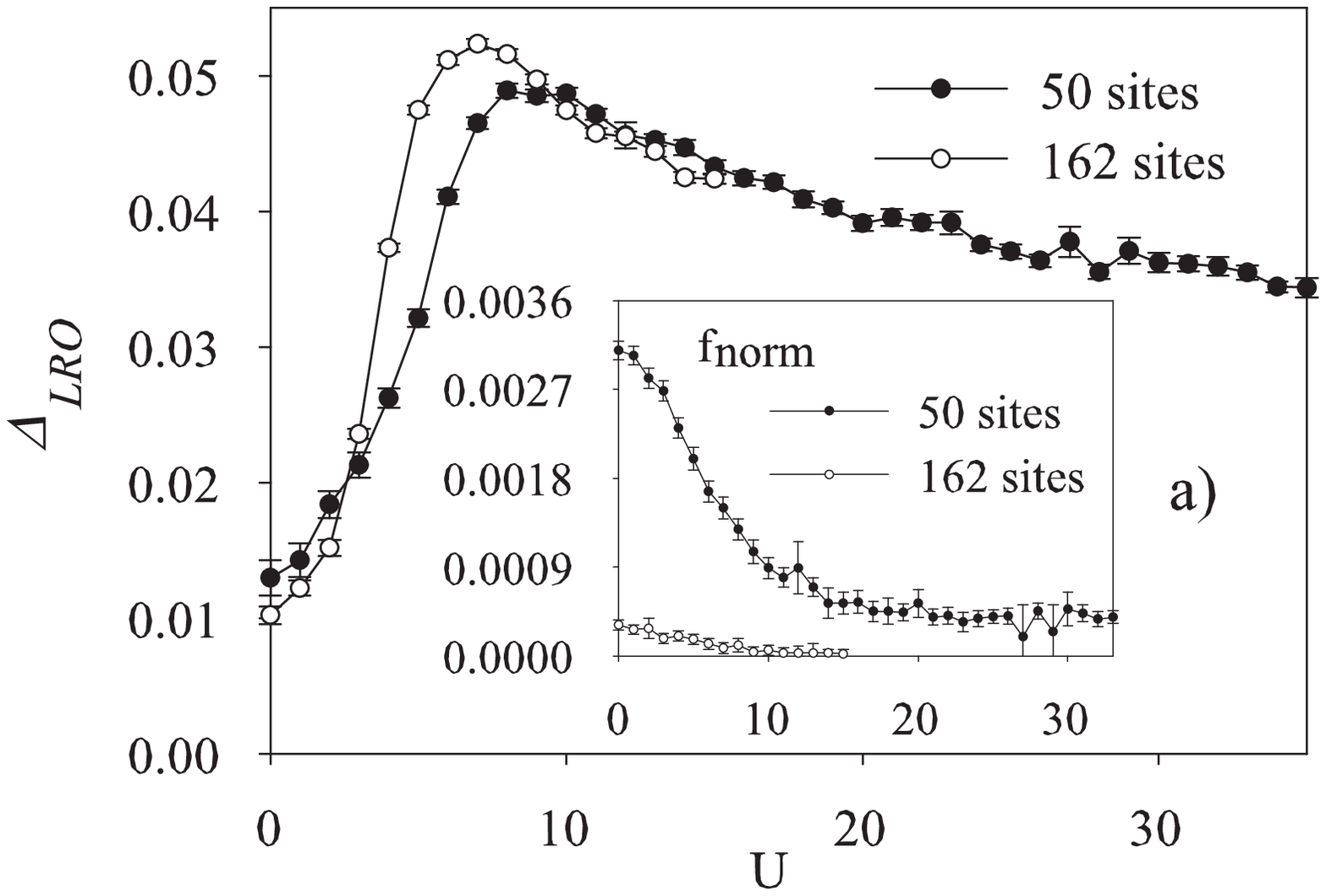}
\includegraphics[width=7.0cm]{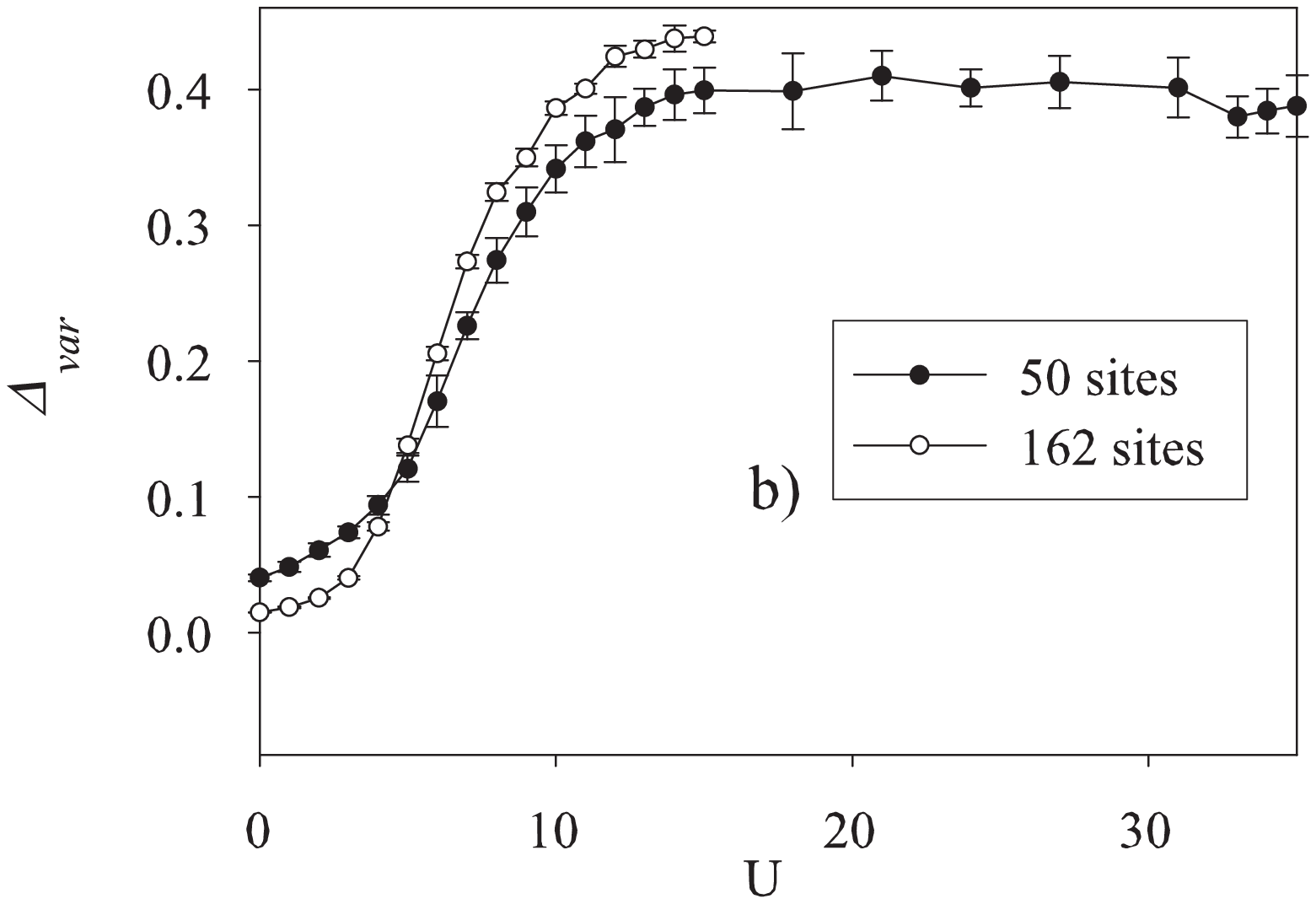}}
\caption{\label{orddel} Superconducting order parameter
$\Delta_{LRO}$ a), and the variational gap $\Delta_{var}$ b), as a
function of $U$ at $V=0$, $J/t=0.2$, $\delta=0.16$. The inset in a) shows the
long distance pairing correlations for the non-superconducting state
(see text).}
\end{figure*}

The behavior of the variational gap $\Delta_{var}$ as shown in
Fig.~\ref{deltas} suggests a crossover from weak to strong
coupling superconductivity as $U$ increases. This is manifest by
comparing Fig.~\ref{orddel}a with  Fig.~\ref{orddel}b. In the latter
the variational energy
gap is displayed for several  $U$'s,
while in the former
we plot the true long range order parameter $\Delta_{LRO}$
 in the correlated wave function.
$\Delta_{LRO}$ is estimated on a finite cluster  through the
 pair-pair correlation function $f$ as follows:
$$
\Delta_{LRO}= \frac{1}{2}\sqrt{f-f_{\rm norm}},
$$
where
\begin{equation}
f\equiv\sum_{\sigma,\sigma^{\prime}}<c^{\dagger}_{\vec{x},\phantom{\vec{1}}\sigma}
c^{\dagger}_{\vec{x}\pm\vec{1},-\sigma}
c^{\phantom{\dagger}}_{\vec{y}\pm\vec{1},-\sigma^{\prime}}
c^{\phantom{\dagger}}_{\vec{y},\phantom{\vec{1}}\sigma^{\prime}}>,
\end{equation}
being evaluated near to the maximum distance $|\vec x - \vec y|$
available on a given size. Notice that $f$ includes normal
contributions $f_{norm}$, which nevertheless vanish in the
 infinite size limit. In order to improve the quality of any
finite size analysis, one should estimate $f_{norm}$ to get a
meaningful   value of the true long range order parameter. We
decided to approximate $f_{norm}$ by the value of $f$ calculated
with the optimized wave function having the same form
(\ref{variational-wf})  with the variational parameter
 $\Delta_{var}$  equal to zero (see
 inset  Fig.\ref{orddel}a). After this subtraction, size
effects are acceptable, at least for our qualitative analysis (see
Fig.\ref{orddel}a).

\begin{figure*}
\centerline{\includegraphics[height=5.5cm]{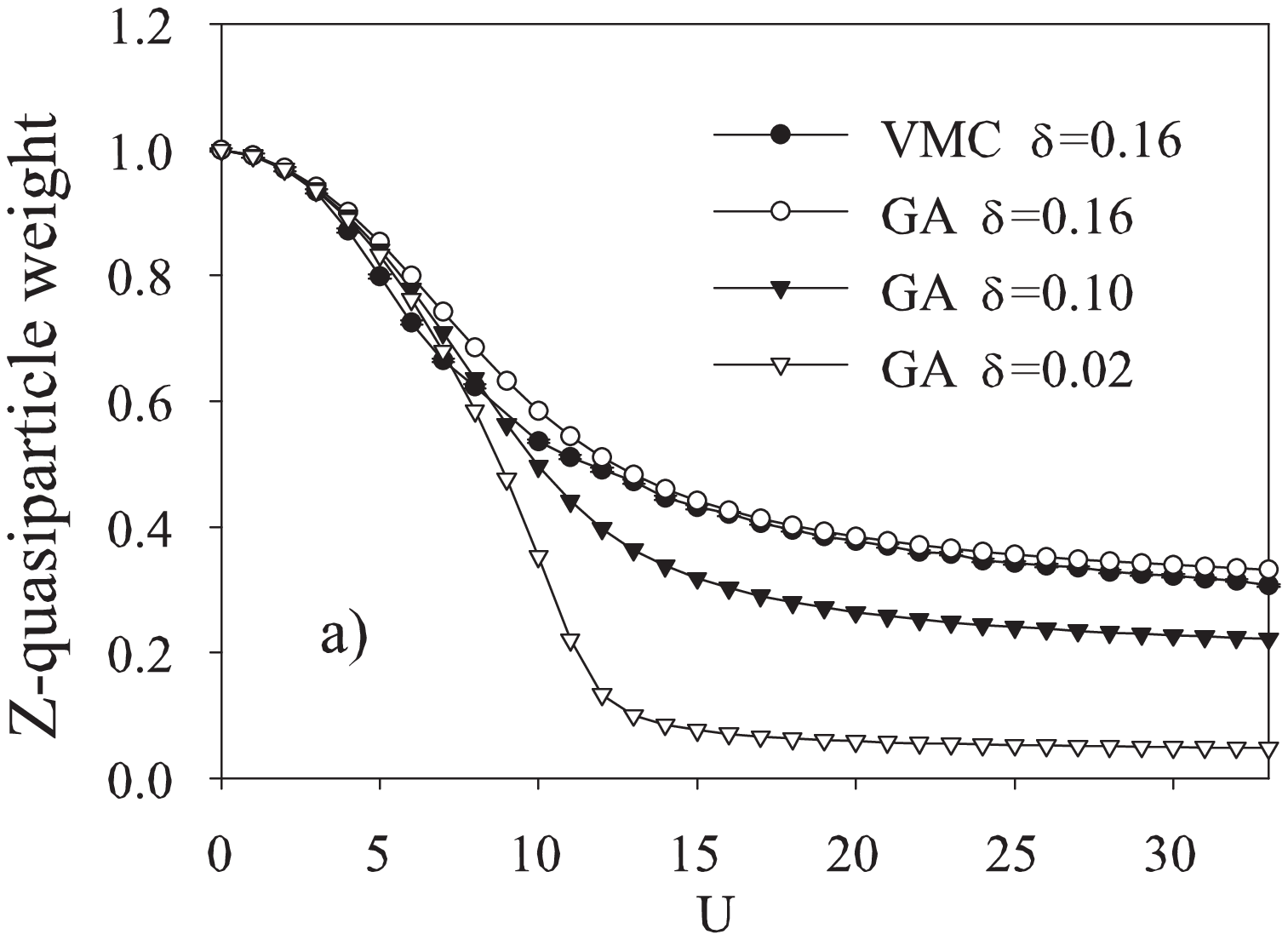}
\includegraphics[height=5.5cm]{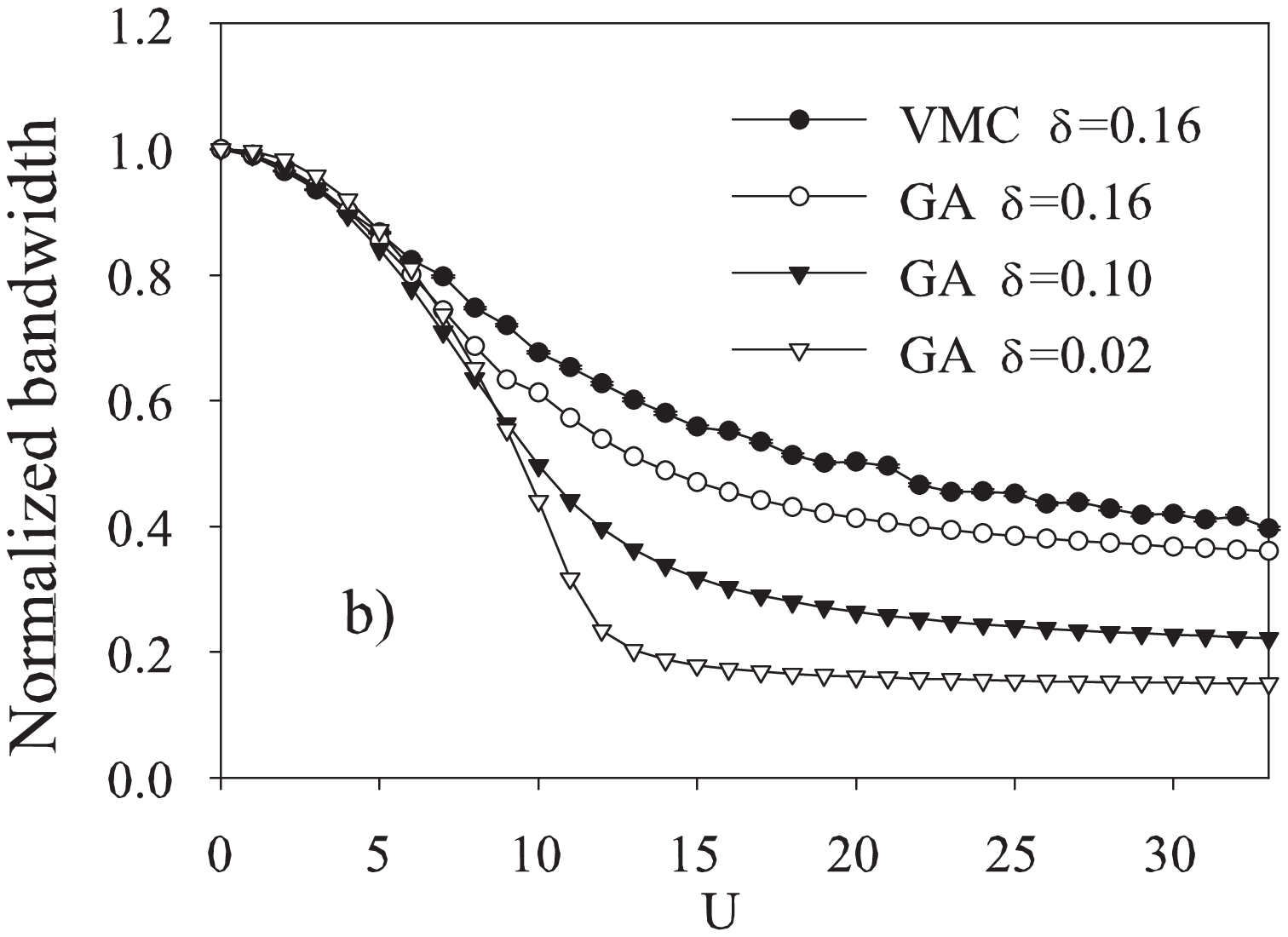}}
 \caption{\label{zband} Panel  a): wave-function
renormalization $Z$ as calculated through the jump in the
momentum distribution along the nodal direction.
Finite  size
scaling  from 50 to 162 is used to evaluate the jump $Z$ in
 the thermodynamic limit for the VMC.
Panel b): quasiparticle bandwidth normalized to its uncorrelated value.
The VMC refers to the 50 site cluster, as finite size effects are small.
 $V=0$ and $J/t=0.2$ for both figures.
 }
\end{figure*}

 The crossover region where
the gap $\Delta_{var}$ rapidly moves from small BCS-like values to much
larger values corresponds to a maximum of the true order
parameter, as one would expect in the intermediate region between
weak to strong coupling superconductivity. The notably difference
with the latter is that in our model the crossover does not occur
by varying the bare attraction $\lambda$, but by increasing
the repulsion $U$.

The different behavior of the variational gap with respect to the
true order parameter, which has been associated with the behavior
of the pseudogap versus T$_c$ in the
cuprates\cite{Sandro,Randeria}, has a clear explanation within
the GA, where $\Delta_{LRO}$ is suppressed by the factor $Z$ with respect
to the uncorrelated $\Delta_{SC}$.
Indeed, as shown in Fig.~\ref{zband}a, the quasiparticle residue $Z$,
defined as the jump in the momentum distribution function along
the nodal directions, is a decreasing function of $U$ tending to
$Z\sim 2\delta$ as $U\to\infty$.

However, as shown in Fig.~\ref{zband}b, $Z$ is not the
reduction factor of the full quasiparticle bandwidth, which gets
contributions also from $J$ and $V$.
Again, this is an
obvious result in the GA where the Hartree-Fock decoupling of the
nearest neighbor interactions effectively generate hopping terms.
In spite of that, the charge current vertex is still determined
by the true hopping $t$, hence gets suppressed by a factor
$Z\simeq 2\delta$. On the contrary, spin current vertex does
include a contribution from $J$ and survives against the strong
wave function renormalization $Z$.

In conclusion, we have shown that strong short range correlations
enhance or suppress pairing correlations if they primarily involve
spin or charge degrees of freedom, respectively. This behavior is
manifest at strong $U$, in agreement with slave boson
approaches\cite{Lee} and numerical
calculations\cite{Gros,Sandro,Randeria}, but starts to appear
already at weak coupling. Indeed, a recent calculation within the
Random Phase Approximation\cite{Plekhanov} shows that the $d$-wave
superconducting phase of model (\ref{Hamiltonian}) at $V=0$ gains
more exchange-correlation energy than a normal metal, thus
supporting the results here obtained by variational Monte Carlo.

{\sl Note added.} When this work was completed we became aware of
a preprint by F.C. Zhang \cite{Zhang} which considers the
Hamiltonian (\ref{Hamiltonian}) with $V=0$ within the GA, in the
context of the gossamer superconductivity scenario recently
proposed by R. Laughlin \cite{Laughlin}. The results of Ref.
\cite{Zhang} qualitatively agrees with our VMC data.

This work was partially supported by MIUR COFIN-2001 and PRA
MALODI.



\begin{thebibliography}{99}
\bibitem{notesdw} We have verified that this variational wavefunction
is stable against a spin-density wave order parameter with  corresponding
magnetic Jastrow factors \cite{shiba},
 in the hole doping region we have considered.
\end{thebibliography}
\end{document}